\documentclass[useAS,usenatbib]{mn2e}
\usepackage{natbib}
\usepackage{graphicx}
\usepackage{multirow}
\usepackage{subfig}
\bibpunct{(}{)}{;}{a}{}{,}
\usepackage{txfonts}
\usepackage{longtable}
\def\ergsec{{\rm ~erg~s^{-1}}}
\def\ergcms{{\rm ~erg~cm^{-2}~s^{-1}}}

\begin{document}
\title[]{Understanding the nature of the intriguing source X Persei: A deep look with a \emph{Suzaku} observation}
%{Intensity Resolved CRSF in X-per}
\author[Maitra et. al.]{Chandreyee Maitra$^{1}$\thanks{Contact e-mail: cmaitra@mpe.mpg.de}\thanks{Present address: Max-Planck-Institut f{\"ur} extraterrestrische Physik, Giessenbachstra{\ss}e, 85748 Garching, Germany},  Harsha Raichur$^{2}$, Pragati Pradhan$^{3}$, Biswajit Paul$^{4}$\\
1.Laboratoire AIM, IRFU/Service d'Astrophysique - CEA/DSM - CNRS - Universite Paris Diderot, Bat. 709, CEA-Saclay, 91191 Gif-sur-Yvette Cedex, France \\
2.~Nordita, KTH Royal Institute of Technology and Stockholm University, Roslagstullsbacken 23, SE-10691 Stockholm, Sweden\\
3.~St Joseph's College, Singamari, Darjeeling 734104, West Bengal, India; North Bengal University, Raja Rammohanpur, District Darjeeling 734013, West Bengal, India \\
4.~Raman Research Institute, C.V. Raman Avenue, Sadashivanagar, Bangalore 560064, India}

\date{Accepted.....; Received .....}

%\pagerange{\pageref{firstpage}--\pageref{lastpage}} \pubyear{2009}

\maketitle

%\label{firstpage}

\begin{abstract}

We present detailed broadband timing and spectral analysis of the persistent, low luminosity and slowly 
spinning pulsar X Persei using a deep Suzaku observation of the source. The spectrum is unusually hard 
with a cyclotron resonance scattering feature (CRSF), the presence of which  has been 
debated. By comparing the spectral models relevant for accretion powered pulsars, we have obtained the best constraint on the broadband spectral 
model of X Persei obtained so 
far. The CRSF is not confirmed in the average spectrum. We have also identified, the presence of different intensity 
levels in the source with distinct changes in the pulse profile and energy spectrum indicating changes 
in the accretion geometry. We further find evidence of a CRSF in the highest intensity levels at $\sim$ 40 keV,
indicating a magnetic field strength of 3.4$\times10^{12}$ G.

\end{abstract}


%\section{INTRODUCTION}
%\section{Observations and Analysis}

\section{Introduction}

X Persei (4U 0352 + 309) is a binary stellar system at a distance of $\sim 0.95$~kpc \citep{Telting.1998} and
consists of a slowly spinning neutron star \cite[$P_{spin} \sim$ 837 s]{White.et.al.1977} which is accreting 
matter from its Be-star companion \citep{Lyubimkov.et.al.1997}. The source is peculiar in that it is a 
persistent source and does not show Type-I outbursts as are commonly observed in other Be-/X-ray binaries. 
But X-ray flares and variability in the X-ray light curve of the source have been observed 
\citep{Lutovinov.et.al.2010,Palombara.et.al.2007}. The orbital period of the binary system is long,
$P_{orb}\sim~250~\rm{d}$, but the orbital eccentricity is only $e \sim 0.11$ \citep{Delgado.2001}. 
The long orbital period and the low X-ray luminosity of the source supports
the assumption that the neutron star in this binary is accreting quasi-spherically by capturing matter from the 
stellar wind of the Be-star \citep{Shakura.et.al.2012}. 
The X-ray spectrum of X Persei is also unlike most accreting X-ray pulsars, the latter can usually be modeled with a power-law and a cutoff in the range 10-30 keV \citep{White.et.al.1983}.
Instead, X Persei has a peculiarly hard X-ray spectrum, making the source detectable at energies higher than 100 keV \citep{Lutovinov.et.al.2010}.
Thus, a single component spectral model is insufficient to model the source continuum spectrum, making it
necessary to include a second high energy component to explain the emission at harder X-rays.
Several authors have used several different models to explain the observed spectrum, of which we note 
the following works. The \emph{RXTE} observations were modelled using a blackbody component at lower energies
and a powerlaw component with an exponential cutoff at higher energies \citep{Coburn.et.al.2001}. A cyclotron resonance scattering 
feature (CRSF) was also required to account for an absorption-like feature seen around 30 keV. The spectrum obtained from \emph{BeppoSAX}
observation was modelled with two powerlaw components with exponential cutoffs at low and high energies
and did not require any CRSF \citep{Salvo.et.al.1998}. The \emph{INTEGRAL} observations used thermal Comptonisation
and bulk motion Comptonization to account for the observed spectrum at the low and high energies respectively \citep{Doro2012}.
The INTEGRAL observations also did not require any CRSF to improve the spectral model. Thus the presence of CRSF noted 
in the \emph{RXTE} observations was not detected in either the \emph{BeppoSAX} or the \emph{INTEGRAL} observation and hence needs to be verified.
There is also no clear agreement on the unusual nature of the X-ray spectrum with respect to the continuum model.

%The X-ray spectrum of X Persei is also unlike most accreting X-ray pulsars, 
%which can be modeled with a power-law and a cutoff in the range 10-30 keV (White, Swank \& Holt 1983). 
%X Persei has an unusually hard X-ray spectrum which can be fit with a two component model comprising 
%two power laws with exponential cutoffs where one dominates at low energies and the other at higher 
%energies (Di Salvo et al. 1998). A hard X-ray tail in the spectrum of X Persei is also reported in 
%some instances (Robba et al. 1996). The \emph{RXTE} spectrum is described by a blackbody at low 
%energies (with kT $\sim$ 1.8 keV ) and a power law with a broad absorption feature $\sim$ 30 keV 
%which has been attributed a CRSF (Coburn et al. 2001). The presence of the CRSF however was not found 
%previously in the BeppoSAX observation (Di Salvo et al. 1998), nor was it confirmed subsequently in the 
%INTEGRAL data (Doroshenko et al. 2012). Hence there is no clear agreement on the unusual nature of the 
%X-ray spectrum of the source, both with respect to the continuum model and the presence of the CRSF.

\begin{figure*}
\includegraphics[angle=-90,scale=0.43]{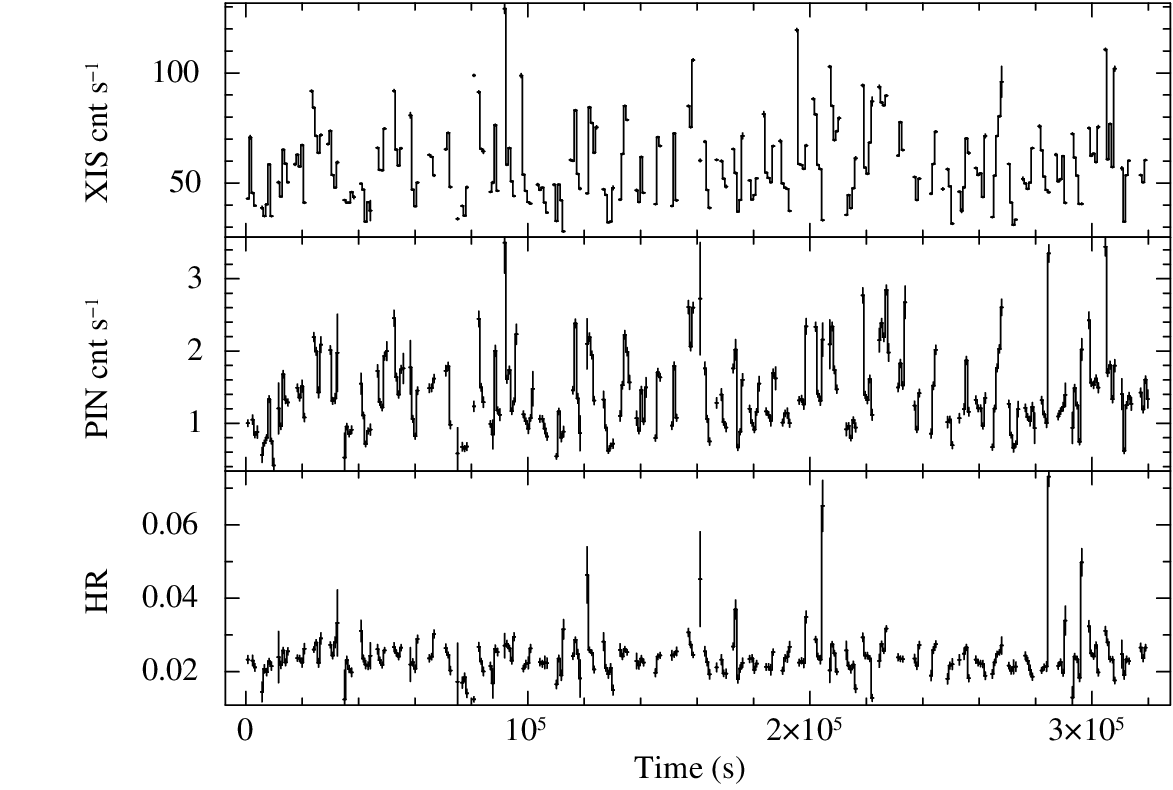}
\includegraphics[angle=-90,scale=0.43]{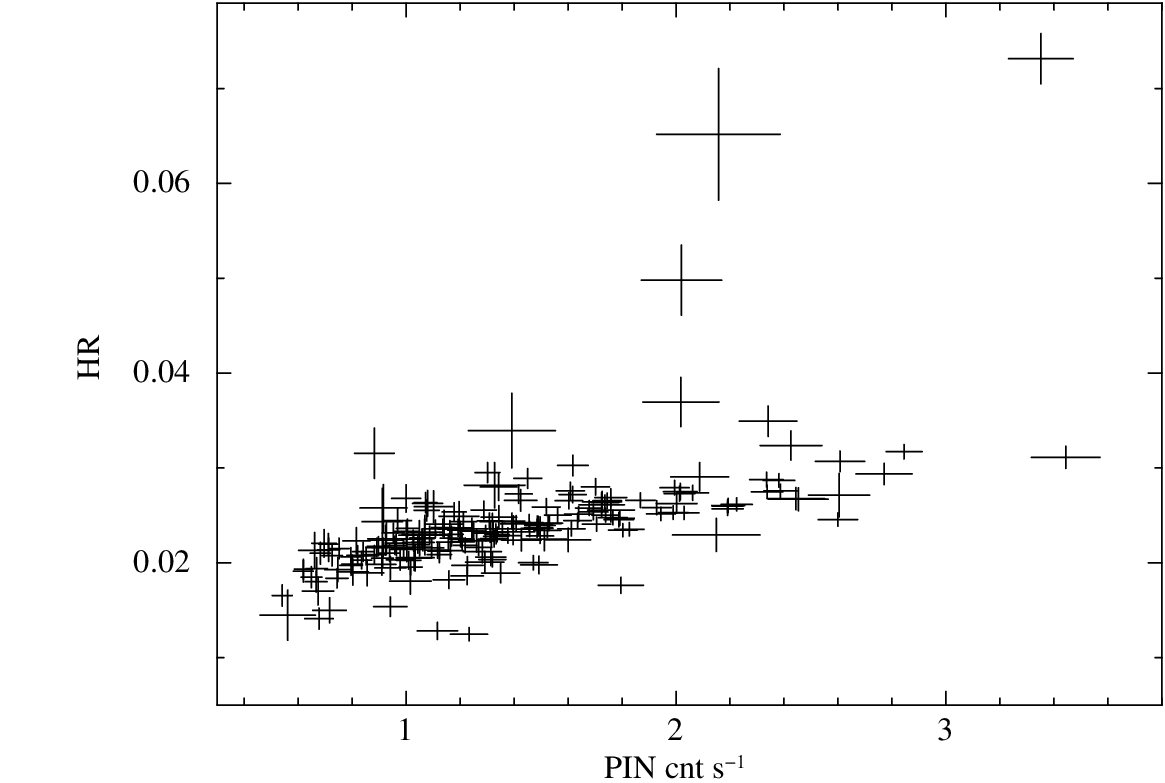}
\caption{Left figure shows the light curves in low and high energy bands 
with the lowest panel showing the hardness ratio. The light curves are binned at the spin period of 835.29 s. 
Right figure shows the HR plotted against the PIN count rate to highlight the variation 
in HR with the count rate in the high energy band. }
\label{HRvsPIN}
\end{figure*}

\section{Observation \& Data Reduction}
X~Persei was observed using \emph{Suzaku} in 2012 for a long duration of 153 ks.
We present here the analysis of this observation. \emph{Suzaku} is a broadband 
X-ray observatory covering an energy range of 0.2-600 keV \citep{Suzaku.07}. The 
two main instruments on board are (i) the X-ray Imaging Spectrometer 
(XIS: covering 0.2-12 keV) (ii) the Hard X-ray Detector (HXD:~PIN diodes covering 
the range of 10-70 keV and GSO crystal scintillators detectors covering 70-600 keV).
Therefore the observation covers a wide energy band simultaneously and with high sensitivity
making it possible to do detailed spectroscopic studies of the source.

The data was reduced and processed with the standard procedures and using the latest CALDB version '20140825'. 
The XIS was operated in the standard data mode with the $\frac{1}{4}$ window mode that provided a timing
resolution of 2 s. The nominal pointing was set to XIS.
XIS data was examined for the presence of pile-up. A central region corresponding to a pile-up fraction 
$>$~4~\% was discarded from all further analysis. 

%The energy spectrum was modeled phenomenologically with different continuum models used to fit the 
%spectra of accretion powered pulsars like the power law with high energy cutoff (highecut, newhcut), 
%or power law with the Fermi Dirac cutoff (fdcut), the cutoff power law (cutoffpl) model, the negative-positive 
%exponential power law component (NPEX), or the thermal comptonization model 'CompTT' and 'COMPMAG' which 
%describes the spectral formation in the accretion column taking into account both thermal and Bulk Motion 
%Comptonization. All the models are available as a standard package in \emph{XSPEC}. 
\section{Timing Analysis: Pulsation search and light curves}
We applied barycentric corrections to the event files for the timing analyis, using the FTOOLS task 'aebarycen'.
Orbital correction was not required as the observation duration is much less than the decoherence timescale due to 
the orbital motion of the system \citep[decoherence timescale is $\sim$ 54 days, Eqn A9]{1997ApJ...474..414C}. 
Light curves were extracted with a time resolution of 2 s and 1 s respectively from the XISs (0.2–12 keV) and HXD/PIN (10–
70 keV). To search for pulsations in the data, pulse folding and $\chi^{2}$ maximization technique was applied. 
Pulsations were found in the light curve at 835.29 $\pm$ 0.29 s, which was subsequently used to bin the light curves and also to create 
corresponding pulse profiles.

Figure \ref{HRvsPIN}, shows the light curves in the low (XIS: 0.2-12 keV) and high (PIN: 10-70 keV) energy bands along 
with the hardness ratio (HR) curve, all binned at the spin period of the source. The 'HR' is defined as the ratio of the count rate
in the XIS band (with the three XISs added together) to the PIN band. It is clear from the 
figure that the count rates in both the energy bands vary by a factor up to 5 during the observation. Moreover the 
HR shows a variation by a similar factor with the PIN count rate (see right of Figure \ref{HRvsPIN}) which 
indicates spectral variability in the source. In order to investigate the true energy dependence of the pulse profiles in this source
and to construct an average model of the representative X-ray 
spectrum of the source, it is important to choose a time interval free 
from intensity or spectral variations. Therefore in order to construct
the pulse profiles and the energy spectrum of X~Persei, we choose a stretch of the observation where 
$0.02 < \rm{HR} < 0.03$ and $ 0.8 < \rm{PIN~c/s} < 1.4$, as not much spectral variation is expected in this narrow stretch. 

\subsection{Energy dependence of the pulse profiles}
\begin{figure*}
\includegraphics[scale=0.53, angle=-90]{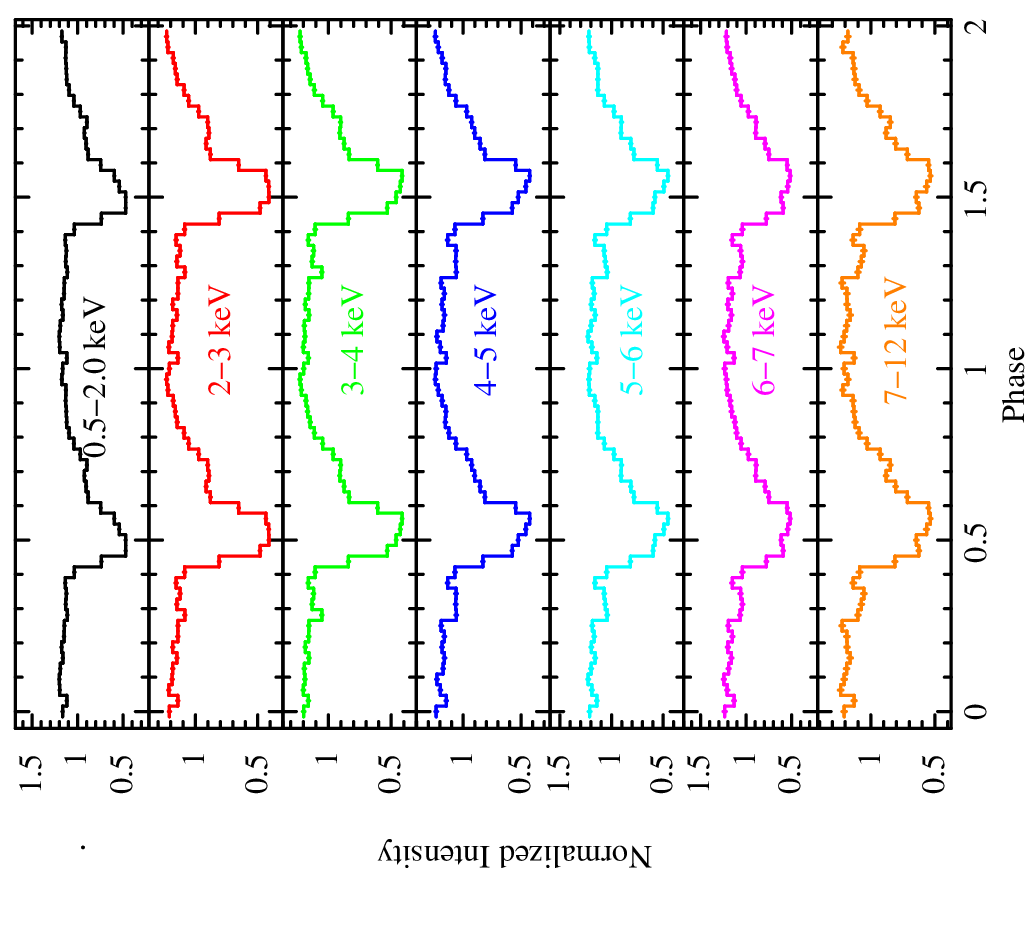}
\includegraphics[scale=0.53,angle=-90]{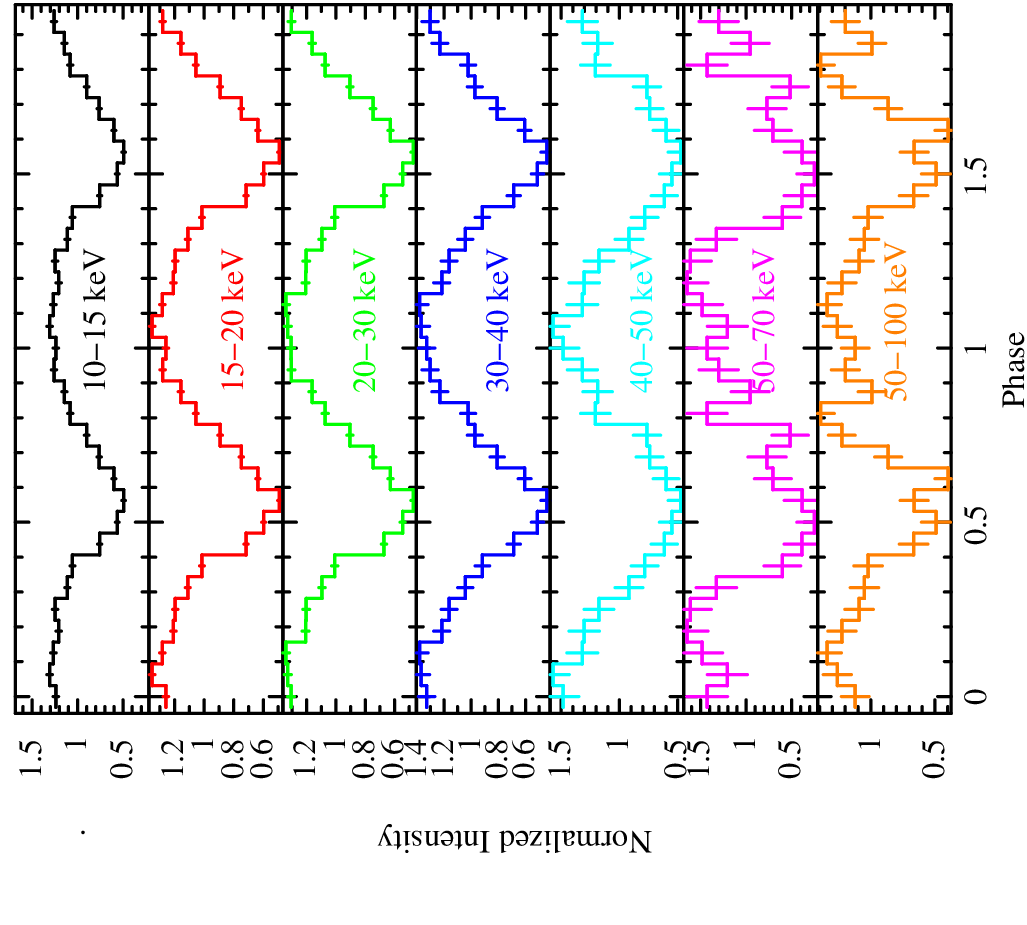}
\caption{Energy-resolved pulse profiles of X~Persei from XIS (left), PIN (right) and GSO data (last figure in the right panel).
The energy ranges are marked inside the panels.}
\label{pp_er}
\end{figure*}

To investigate any possible energy dependence of the pulse profiles, we created 
the energy resolved pulse profiles by folding the data in different energy bands at the obtained pulse period.
We have only used XIS and PIN light curves for the purpose between the energy ranges 0.5--12 and 10--70 keV respectively. Pulsations are detected up to 
100 keV asserting the hard nature of the source as seen from the pulse profile created using the GSO light curve. 
Energy dependent pulse profiles are shown in figure \ref{pp_er}. 
The pulse profiles are more or less sinusoidal in shape from 0.5--100 keV, with little evolution with energy. This is consistent
with previous results from \emph{RXTE} and \emph{BeppoSAX} observations \citep{Salvo.et.al.1998,Coburn.et.al.2001}. 
An additional spike like feature is evident at 
phase $\sim$ 0.4 in the low energy pulse profiles ($\le 12$ keV) which do not show any evolution of its strength with energy in these
energy bands. The high energy pulse profiles are featureless.

\section{Spectral Analysis}
Spectral analysis was performed using \textit{XSPEC} v12.8.1 \footnote{https://heasarc.gsfc.nasa.gov/xanadu/xspec/}.
The XIS spectra were fitted from 0.8-10 keV and the PIN spectrum from 10-70 keV and GSO between 50-110 keV.
The energy range of 1.75-2.23 keV was neglected due to artificial structures in the XIS
spectra around the Si edge and Au edge. For XIS spectrum, we grouped the energy ranges between
0.8-6 keV and 6--7 keV by  8, 4 and the rest by a factor of 8 respectively. This ensured that we had enough counts in each channel
to apply $\chi^{2}$ statistics for fitting the spectra. The energy around Fe line (6--7 keV) was binned by a smaller factor to avoid loosing information
on a narrow line feature, if any. The whole energy range 
of PIN spectrum was binned by a factor of 4, for the reason mentioned above. For binning the GSO spectrum, we used the suggested grouping method for GSO spectrum with `gsobgd64bins.dat' available 
in the \emph{Suzaku} analysis homepage\footnote{http://heasarc.gsfc.nasa.gov/docs/suzaku/analysis/abc/}.

\subsection{Broad band spectroscopy and the average energy spectrum}\label{avspec}

The spectra of accretion powered pulsars are usually modelled phenomenologically using a powerlaw with quasi exponential high energy cutoff 
of various functional forms. The most widely used models are a power-law with a high energy
exponential cutoff, a broken power-law \citep{white1983, mihara1995, coburn2001}, or a Fermi-Dirac cutoff \citep[{\it XSPEC} model 'fdcut']{tanaka1986}. More recently, 
a smoothed high energy cutoff model ({\it XSPEC} model 'newhcut') is also used extensively as it avoids artificial residuals in the spectra
due to the abrupt nature of the exponential cutoff model. Other models are a combination of two powerlaws with different photon
indices but common cutoff energy value called the Negative and Positive powerlaws with Exponential model \citep[{\it XSPEC} model 
'NPEX']{mihara1995,makishima1999}, and a thermal Comptonization model ({\it XSPEC} model 'CompTT model). Recently, a new model for X-ray continuum formation has been developed for accretion powered
pulsars with high magnetic fields ($B\ge10^{12}$\,G) assuming a cylindrical geometry and a constant B permeating it. In this model, the source
of seed photons are assumed to be from the blackbody photons arising from the base of the accretion column. This model was also successfully applied to the spectra
of Supergiant Fast X-ray Transients which are relatively low luminosity sources ($L_{\rm x} \le 10^{34}$\,$\ergsec$), and is available as a standard
{\it XSPEC} model 'COMPMAG' \citep{F12a}. The model allows fitting parameters related to the physics of accretion on to a high magnetic field
neutron star, for example $\eta$ (index of the velocity profile), $\beta$ (terminal velocity of the accreting matter at the neutron star surface), $r_{0}$ (radiation
of the accretion column in terms of Schwarzchild radius) in addition to the electron temperature and optical depth of the accreting column and 
temperature of the seed blackbody photon.
An updated version of this model ({\it XSPEC} model 'COMPMAG2') has also been developed by the authors which also considers
the contribution of Bremmstrahlung and cyclotron emission as the source of seed photons for Comptonization \citep{F16}. This effect is more 
relevant for high
luminosity X-ray pulsars where the contribution of Comptonized blackbody has been found to be negligible with respect to the Comptonized Bremmstrahlung
and cylotron emission. The model also has many additional parameters compared to 'COMPMAG', for example the height of the accretion column in addition to it's radius,
and the magnetic field strength B
of the neutron star surface. Additionally it provides options to choose the source of seed blackbody photons, as well as the beam pattern 
between pencil and fan.

We tried to fit the broadband energy spectrum of X~Persei with all the different continuum models mentioned above. 
The spectrum extends up to 100 keV
and as in the previous works using the BeppoSAX \citep{Salvo.et.al.1998}, 
\emph{RXTE} \citep{Coburn.et.al.2001}, and \emph{INTEGRAL} \citep{Doro2012} observations, a two component 
model is required to fit separately the low and the high energy part of the broadband \emph{Suzaku} 
spectrum (using XIS+PIN+GSO). Spectra obtained from XIS, PIN and GSO instruments
were fitted simultaneously allowing for cross-calibration constants for the PIN and GSO instruments with 
respect to the XIS instrument. In all the cases we applied a Galactic line of sight absorption (\textit{XSPEC} model 'tbabs') component
to the continuum models. No Fe line was detected in the spectrum. A broad emission feature at 1.65 keV was however detected only in the XIS1 (BI CCD) spectrum.
As this was detected only in one of the three XISs this may be an instrumental feature and was hence not included in the analysis.

For all the phenomenological continuum models, 
a broad absorption feature ($\sim$ 30 keV) was evident in the 
residuals, reminiscent of the CRSF reported from \emph{RXTE} observations. Adding CRSF absorption 
profile (`cyclabs' in \emph{XSPEC}) to the spectrum improved the fit. 
Another point we noted is that since the spectrum of 
X~Persei is unusually hard and extended $>$ 100 keV, the GSO spectrum was crucial to constrain the cutoff in the 
high energy range, as it mostly appeared to be a hard tail like feature in the PIN band (10-70 keV). Overall, we 
found a two component power law model with smooth exponential rollover (model 'newhcut') as the best representative phenomenological continnum
model of the 
X-ray spectrum of X Persei, since it provided the best-fit quality with respect to the reduced $\chi^{2}$ (1.70 for 567 dof). 
The corresponding decrease in $\chi^{2}$ after the inclusion of the CRSF was 11. However,
the observed feature was shallow and broad, with a width of $\sim$ 12 keV. This is broader than most of the CRSFs observed, and a broad
absorption feauture as this could artificially modify the continnum spectrum considerabl. Therefore, it could arise from an artifact due to inadequate
modelling of the continuum emission, and should be treated with caution.
The best-fit continuum parameters obtained using the 'newhcut' model are consistent with the results obtained using BeppoSAX observation. 
However,  \emph{Suzaku} being more sensitive than the former at the high energy range, the high energy parameters 
are better constrained. 

\begin{figure}
\hspace{-0.5cm}
\includegraphics[scale=0.35,angle=-90]{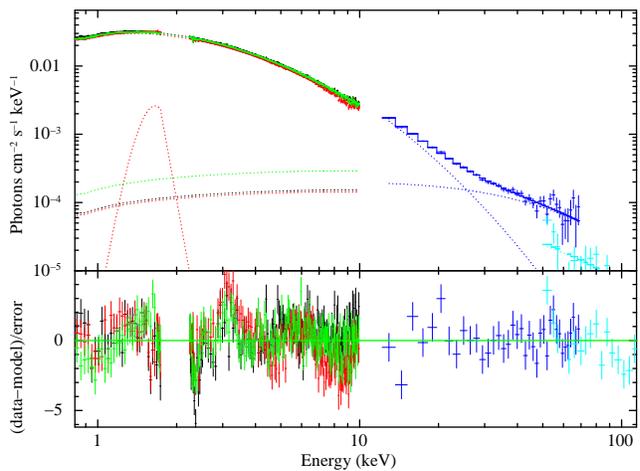}
\caption{The upper panel shows the unfolded best fit average (or representative) energy spectrum
of X~Persei showing the different continuum model components; the bottom panel shows the residues
of the best fit model.}
\label{BF_avg_spec}
\end{figure}

Thereafter, we proceeded to fit the broadband spectrum with the analytical
model 'COMPMAG' and its updated version. The 'COMPMAG' model has many spectral parameters which were impossible to constrain simultaneously.
Especially, $\beta$ if left free preferred a value $>$ 0.9 in the fit, which is
greater than the maximum possible terminal velocity at the neutron star surface.
Therefore we tested the two component 'COMPMAG' model, assuming reasonable parameter values applicable to low luminosity accretion powered
pulsars, {\it i.e.} free-fall velocity profile ($\eta$=0.5),
and $\beta$=0.5 which corresponds to the maximum terminal velocity at the neutron star surface, and a pencil beam emission pattern.
The radius of the accretion column ($r_{0}$) was set to 0.25 corresponding to a radius of $\sim$ 1 km 
for a neutron star of mass 1.4\,M$_{\odot}$. The Albedo at the neutron star surface (A) was set to 1. This provided an acceptable fit to the data
with acceptable and physically viable parameter values. Addition of a CRSF feature was not required for this model.
The high $\beta$ preferred by the fit indicated that Bulk Motion Comptonization (BMC)
dominates the Comptonization process in X~Persei. The fit also provided a low value of electron temperature ($kT$) consistent with the high value of $\beta$
obtained \citep{F12a}. The dominance of BMC in low luminosity pulsars, especially X~-Persei has been predicted before and is consistent with our 
results \citep{wolff2005,wolff2007}. In \cite{wolff2007}, the authors qualitatively described the observed spectrum of X~-Persei with a pure BMC
model. The obtained reduced $\chi^{2}$ value with 'COMPMAG' was higher than that obtained with the 'newhcut' model (1.84 for 569 dof), although there were
no systematic pattern in the residuals, and the higher $\chi^{2}$ was mainly contributed by a higher variance in a few energy bins
of the XIS spectra. Fit with the updated 'COMPMAG2' model was difficult with the given statistical quality of the data,
as it required fitting many more additional parameters. Moreover as discussed in \cite{F16}, the 'COMPMAG' model provides an adequate
description of the spectrum for low luminosity pulsars, where the blackbody emission provides the major source of photons for Comptonization, 
appropriate for the case of X~Persei. Considering the above, and as 'COMPMAG' is a more physical model to understand
the continuum spectra of the source, we used this as our best-fit model for the rest of the paper. As a CRSF feature is not required
for this model, we cannot claim the presence of a CRSF in the average spectrum of X~Persei.
Table \ref{tab-spec} summarizes the best-fit broadband spectral parameters obtained using the 
`newhcut' (without addition of the CRSF) and 'COMPMAG' model. Fig \ref{BF_avg_spec} shows the best-fit unfolded 
spectrum for the best fit model, showing the model components. 

\begin{table}
\caption{Best fitting phase averaged spectral parameters of X~persei. Errors quoted are for 90 per cent confidence range.}
\begin{tabular}{c c c}
\hline \hline

parameters & NEWHCUT & COMPMAG \\
\hline
$\ensuremath{N_{\mathrm{H}}}$ ($10^{22}$ atoms $cm^{-2}$) & 0.39 $\pm$ 0.003 & $0.17 \pm$ 0.005\\
$\Gamma_1$ (low energy) & 0.28 $\pm$ 0.01 & --\\
%norm_1 & $(2.5 \pm 0.3) \times 10^{-2}$ & $(2.21 \pm 0.3)\times 10^{-2}$ & $(2.3 \pm 0.2)\times 10^{-2}$ & $(1.6 \pm 0.3)\times 10^{-2}$ \\
E1-folding energy (keV) &  $3.49 \pm 0.03$ & --\\
E1-cut energy (keV) & $2.98 \pm 0.03$ & --\\
$\Gamma_2$ (high energy) & $1.52 \pm 0.03$ & --\\
E2-cut energy (keV) & $56.26 \pm 10$ & --\\
$KT_{bb}$ (keV)  & -- & $0.86 \pm 0.03$\\
$kT$1 (keV) & -- & $0.21 \pm 0.02$\\
$\tau$1  & -- & $0.12 \pm 0.02$ \\
$kT$2 (keV) & -- & $2.3 \pm 1.2$\\
$\tau$2  & -- & $6.3 \pm 0.5$ \\
Flux (XIS) $^a$ (0.3-10 keV) & $ 6.62\times10^{-10}\pm 0.41 $ & $6.75\times10^{-10}\pm 0.43 $\\
Flux (PIN) $^a$ (10-70 keV) & $7.43\times10^{-10} \pm 0.11$ & $7.40\times10^{-10} \pm 0.12$\\
$L_{{\rm x}}^b$ (0.3-70 keV) & $1.6\times10^{35}$ & $1.6\times10^{35}$\\
reduced $\chi^{2}$/d.o.f & 1.73/569 & 1.83/569\\
\hline
\end{tabular}\\
$^{a}$ - Observed flux in units of $\ergcms$ in the mentioned energy band.\\
$^{b}$ - Bolometric unabsorbed luminosity in units of $\ergsec$ in the mentioned energy band, assuming a distance of 1 kpc.\\
\label{tab-spec}
\end{table}

\begin{figure}
\includegraphics[scale=0.45,angle=-90]{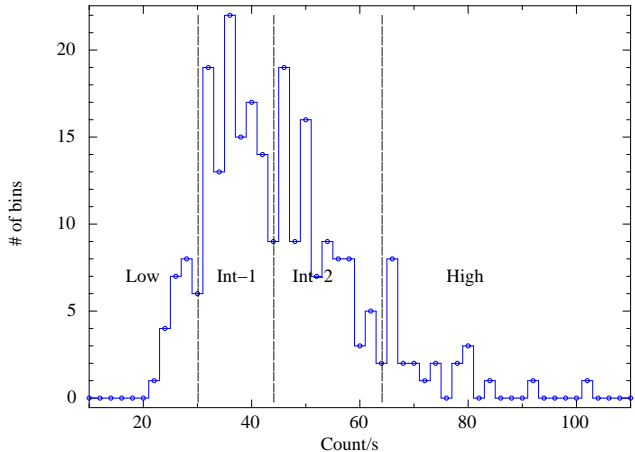}
\caption{Figure shows the histogram of the combined light curve of X~Persei 
as observed by the \emph{Suzaku} XIS detectors. Count rates demarking different intensity levels used in the subsequent analysis are 
shown with vertical lines
}
\label{hist}
\end{figure}

\section{Intensity levels}
 \label{int}
The light curves of X~Persei in both the low and high energy bands show variations in the count rate up to a factor of 5. 
As mentioned earlier, large variation in the 'HR' is also observed (see Figure \ref{HRvsPIN}). This motivated us to
search for different
intensity levels in the observation. This was done using the combined XIS light curve. 
For this, the light curve was first binned using the
determined spin period of the source. Then we plotted the number of bins as a function of the count rate to get a 
histogram which is shown in Figure \ref{hist}. 
From the histogram, we identified the 'High' and 'Low' levels from the two ends of the distribution, or the tail regions. 
The intermediate level was identified from the peak of the distribution. Figure \ref{hist} also hinted a double peaked
feature of the distribution. Hence we subdivided the intermediate level into 'Intermediate1' (Int-1 from now) and
'Intermediate2' (Int-2 from now) levels.

The intensity levels chosen for this work are also shown in the figure, and the details are tabulated in Table 2. \\
%Low state:~Count rate $< 31$~c/s, Int-1 state:~31~c/s~$<$~Count rate~$<$~45~c/s, 
%Int-2 state:~45~c/s~$<$~Count rate~$<$~65~c/s, High state:~Count rate~$>$~65~c/s.
To investigate whether these are distinct intensity levels associated with changes in the accretion geometry 
we investigated the pulse profiles and the broadband spectrum of the source at the different intensities. 
The results are presented in the subsequent subsections. Changes in the pulse profiles and the energy
spectrum with intensity is associated with corresponding changes in the conditions of the accretion column \citep{becker2012}.
\begin{table*}
%\centering
\begin{center}
\resizebox{1.1\textwidth}{!}
 {\begin{minipage}{\textwidth}
\begin{tabular}{ccc}
\hline
Intensity Level & Count rate (c/s) & Exposure (ks)  \\
\hline
\hline
Low & $< 31$ & 11.4 \\
Int-1 & $\ge 31 \& \le 45$ & 54.2 \\
Int-2 & $\ge 45 \& \le 65$ & 65 \\
High & $>65$ & 10.5 \\
\hline
\end{tabular}
\caption{Table detailing the different intensity levels chosen for this work.}
 \end{minipage}}
 \end{center}
 \label{int-table}
\end{table*}

\begin{figure*}
\hspace{-2.0cm}
\includegraphics[scale=0.32,angle=-90]{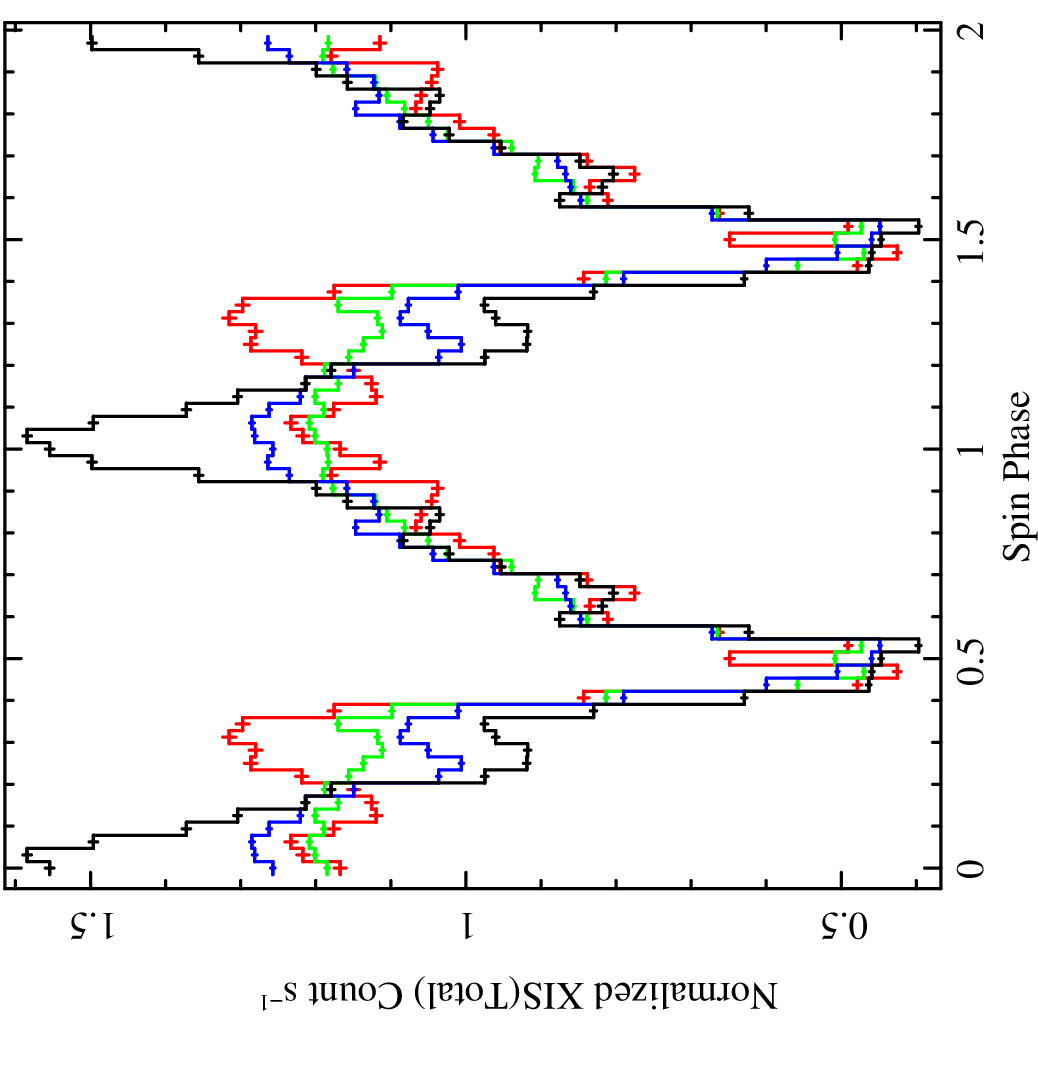}
\includegraphics[scale=0.32,angle=-90]{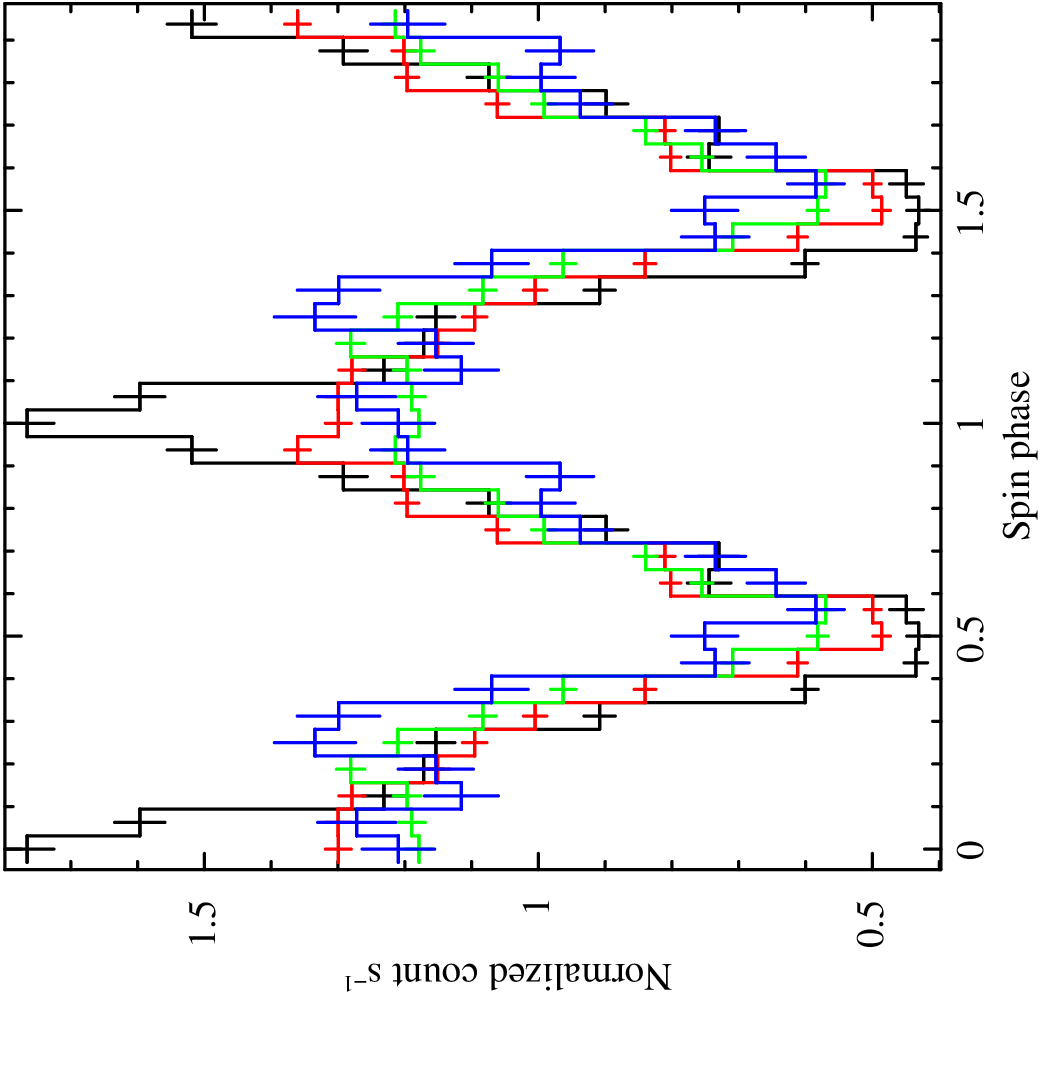}
\includegraphics[scale=0.32,angle=-90]{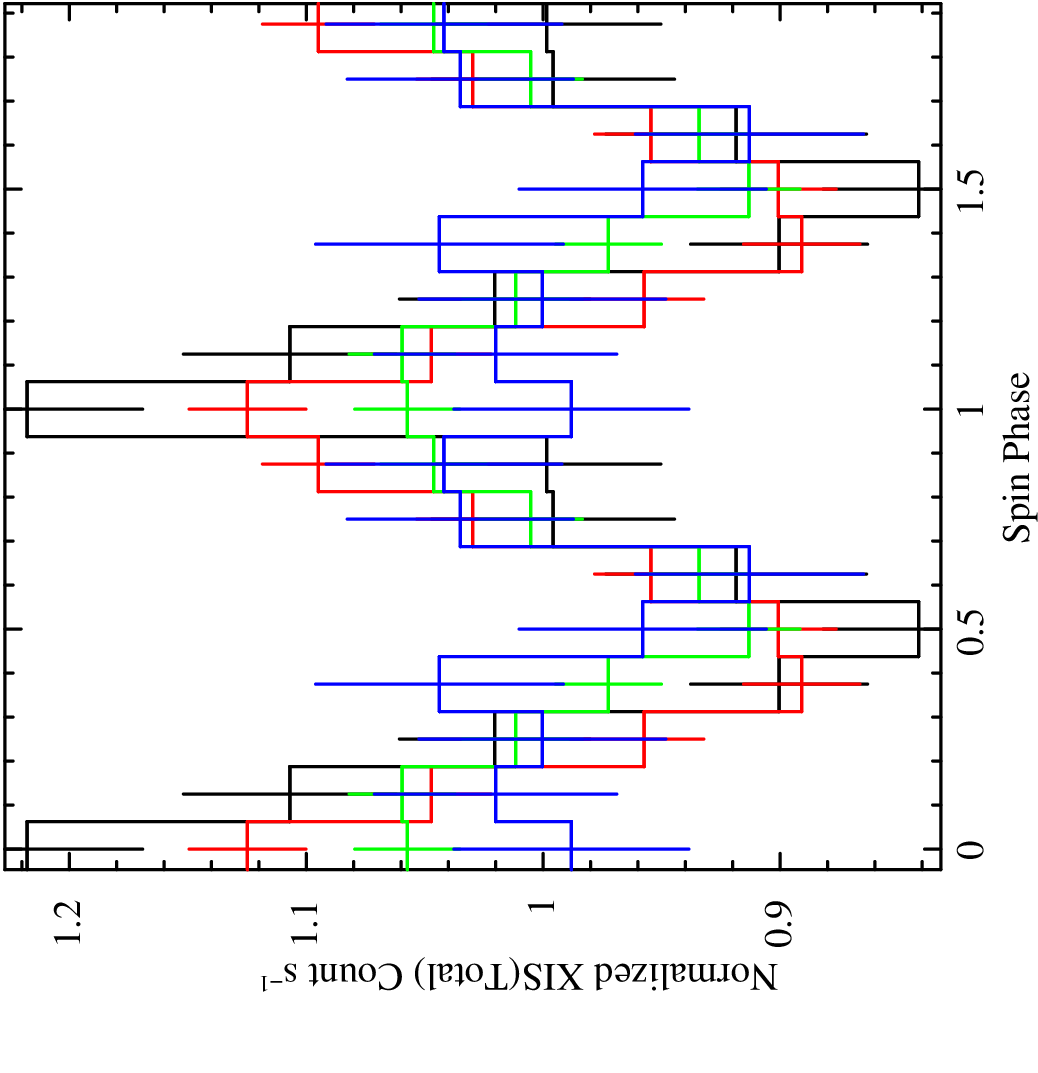}
\caption{Normalised background subtracted pulse profiles of X~Persei at different intensity levels constructed 
from XIS (0.5-10 keV) (left panel), PIN (10-70 keV) (middle panel) and GSO (50-100 keV) (right panel) light curves respectively.
Pulse profiles in Low, Int-1, Int-2 and High intensity levels are plotted in red, green, blue and black respectively. }
\label{pp}
\end{figure*}

\begin{table*}
\caption{Best fitting spectral parameters of X~persei for different intensity levels. Errors quoted are for 90 per cent confidence range.}
\begin{tabular}{c c c c c }
\hline \hline

parameters & High & Int-2 & Int-1 & Low \\
\hline
$\ensuremath{N_{\mathrm{H}}}$ ($10^{22}$ atoms $cm^{-2}$) & 0.20 $\pm$ 0.008 & 0.20 $\pm$ 0.002 & 0.17 $\pm$ 0.005 & 0.14 $\pm$ 0.009  \\

$KT_{bb}$ (keV)  & $0.95 \pm 0.02$ & $0.90 \pm 0.01$ & $0.86 \pm 0.03$ & $0.83 \pm 0.05$\\
$kT$1 (keV) & $0.23 \pm 0.10$ & $0.22 \pm 0.02$ & $0.20 \pm 0.05$ & $0.20 \pm 0.15$\\
$\tau$1  & $0.12 \pm 0.03$ & $0.11 \pm 0.01$ & $0.10 \pm 0.02$ & $0.10 \pm 0.04$ \\
$kT$2 (keV) & $2.8 \pm 0.5$ & $2.3 \pm 0.2$ & $2.3 \pm 1.1$ & $1.7_{-0.5}^{+2.2}$\\
$\tau$2  & $6.0 \pm 0.2$ & $6.0 \pm 0.1$ & $6.0 \pm 0.4$ & $7.0 \pm 0.7$ \\
\\
CRSF $D$ & $0.54 \pm 0.15$ & $0.22_{-0.08}^{+0.13}$ & --  & --\\
CRSF $E$ (keV) & $37.8 \pm 2.4$ & $40.8 \pm 2.0$ & -- & --\\
CRSF $W$ (keV) & $5.8_{-2.4}^{+2.0}$ & $3.7_{-2.1}^{+5.2}$ & -- & --\\ 
\\
CRSF $D_1$ (keV) & -- & 0.15 $\pm$ 0.08 & --  & --\\
CRSF $E_1$ (keV) & -- & 28.6 $\pm$ 1.5 & -- & --\\
CRSF $W_1$ (keV) & -- & 3 (fixed) & -- & --\\ 
Fe K$\alpha$ (keV) & -- & -- & $6.40\pm$ & --\\
eqw (eV) & -- & -- & $7.5\pm2.0$ & --\\
Flux (XIS) $^a$ (0.3-10 keV) & $ 11.97\times10^{-10}\pm 0.03 $ & $9.18\times10^{-10}\pm 0.02$ & $9.76\times10^{-10}\pm 0.01$ & $4.91\times10^{-10}\pm 0.02$ \\
Flux (PIN) $^a$ (10-70 keV) & $14.12\times10^{-10} \pm 0.91$ & $10.73\times10^{-10} \pm 0.24$ & $8.32\times10^{-10} \pm 0.71$ & $3.69\times10^{-10} \pm 0.65$\\
$L_{{\rm x}}^b$ (0.3-70 keV) & $3.0\times10^{35}$ & $2.3\times10^{35}$ & $1.6\times10^{35}$ & $1.4\times10^{35}$\\
reduced $\chi^{2}$/d.o.f (no CRSF) & -- & -- & 1.87/557 & 1.21/555\\
reduced $\chi^{2}$/d.o.f (1 CRSF) & 1.35/573 & -- & -- & --\\
reduced $\chi^{2}$/d.o.f (2 CRSFs) & -- & 1.95/554 & -- & --\\
\hline
\end{tabular}\\
$^{a}$ - Observed flux in units of $\ergcms$\\
$^{b}$ - Bolometric unabsorbed luminosity in units of $\ergsec$ in the mentioned energy band, assuming a distance of 1 kpc.\\
\label{avg_spec_params_intensity}
\end{table*}

\begin{figure*}
\includegraphics[scale=0.3,angle=-90]{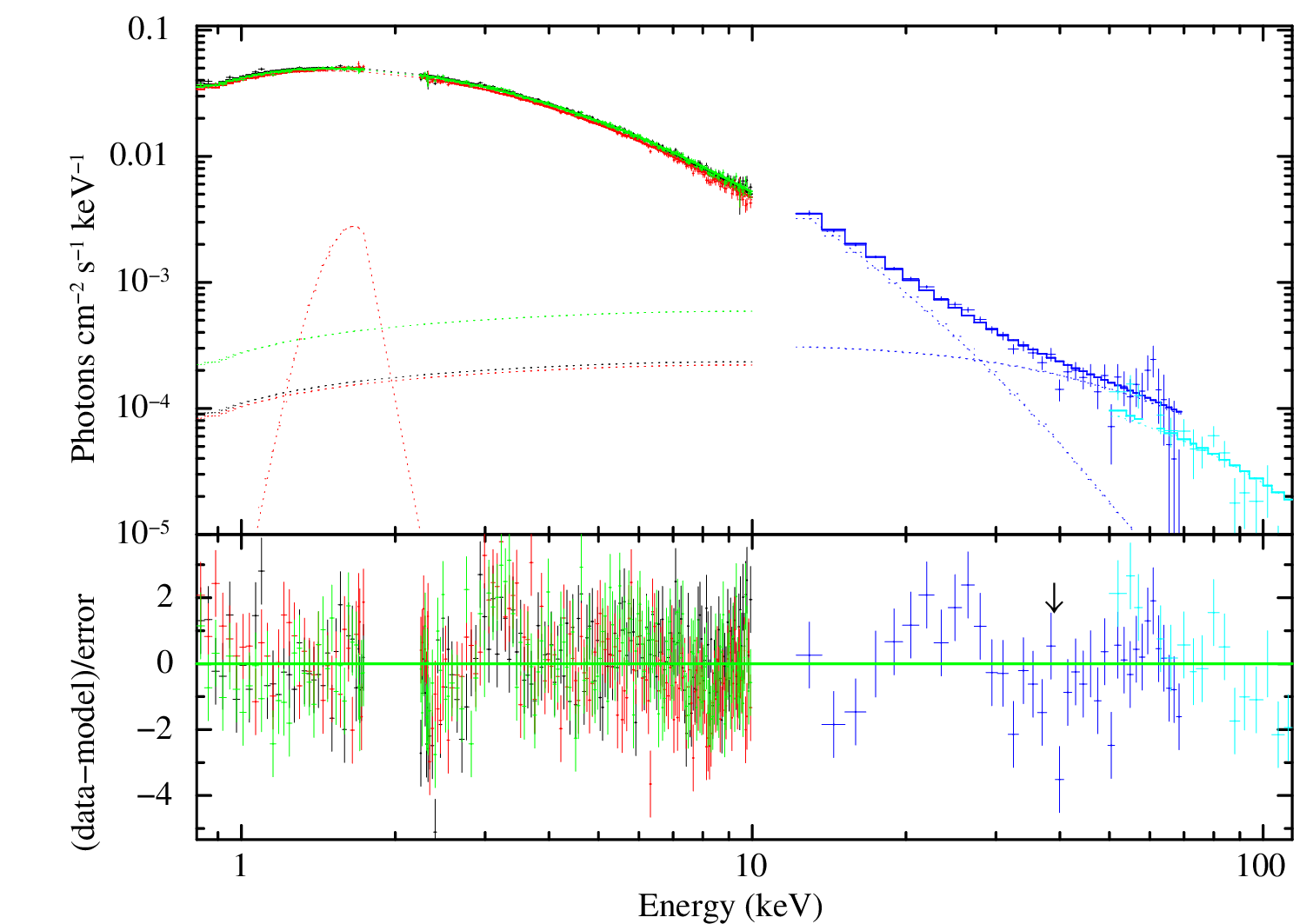}
\includegraphics[scale=0.3,angle=-90]{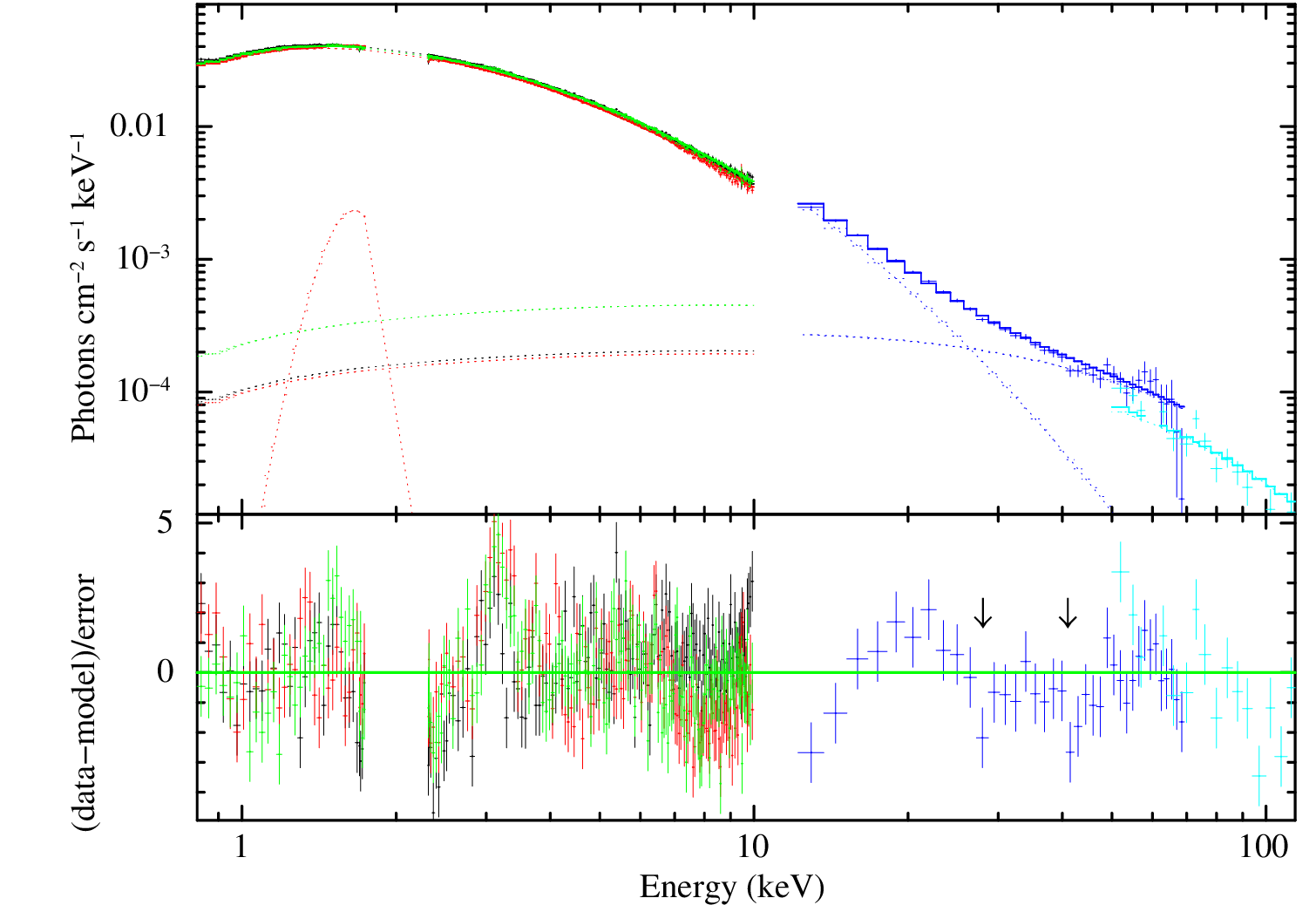}
\includegraphics[scale=0.3,angle=-90]{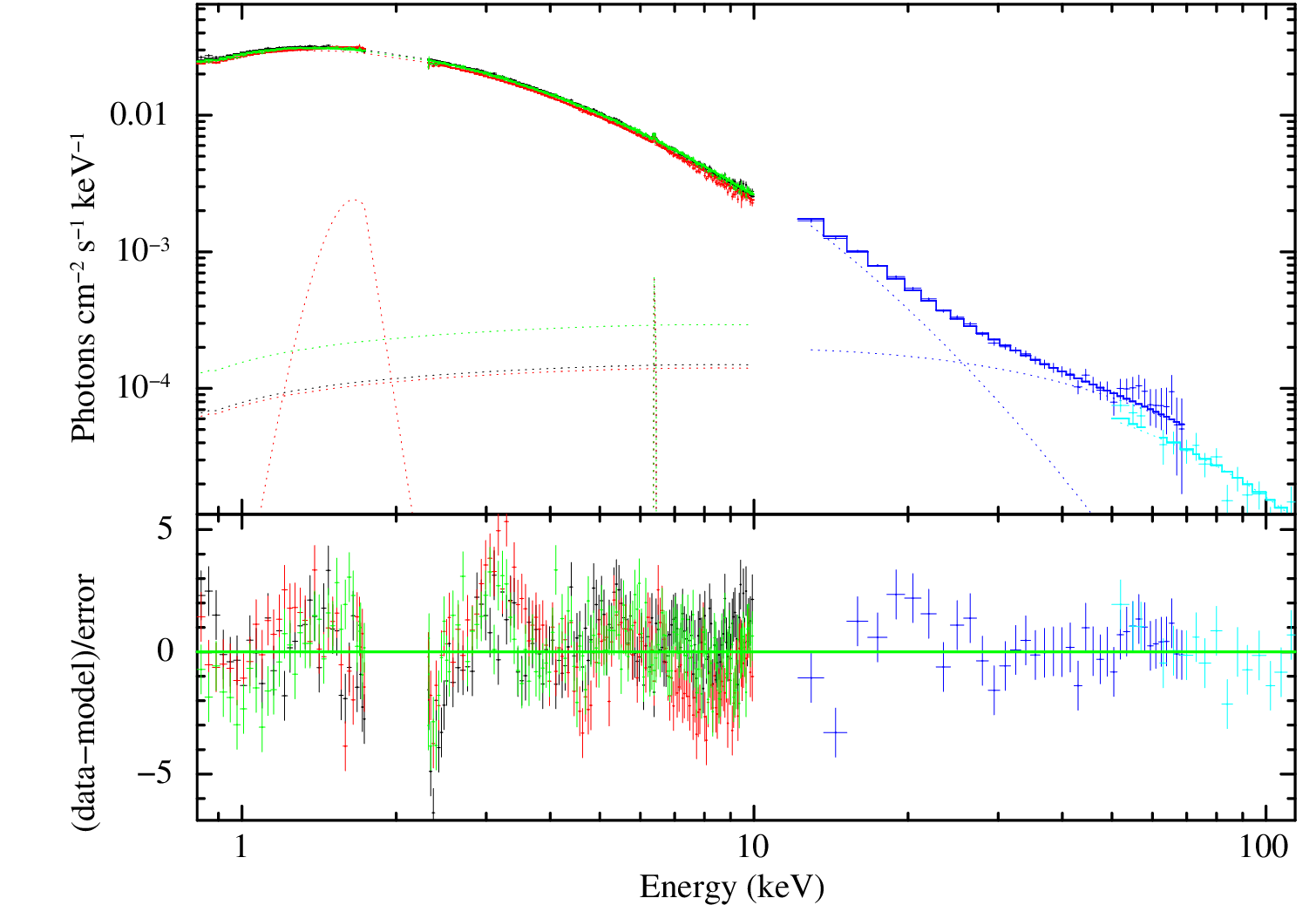}
\includegraphics[scale=0.3,angle=-90]{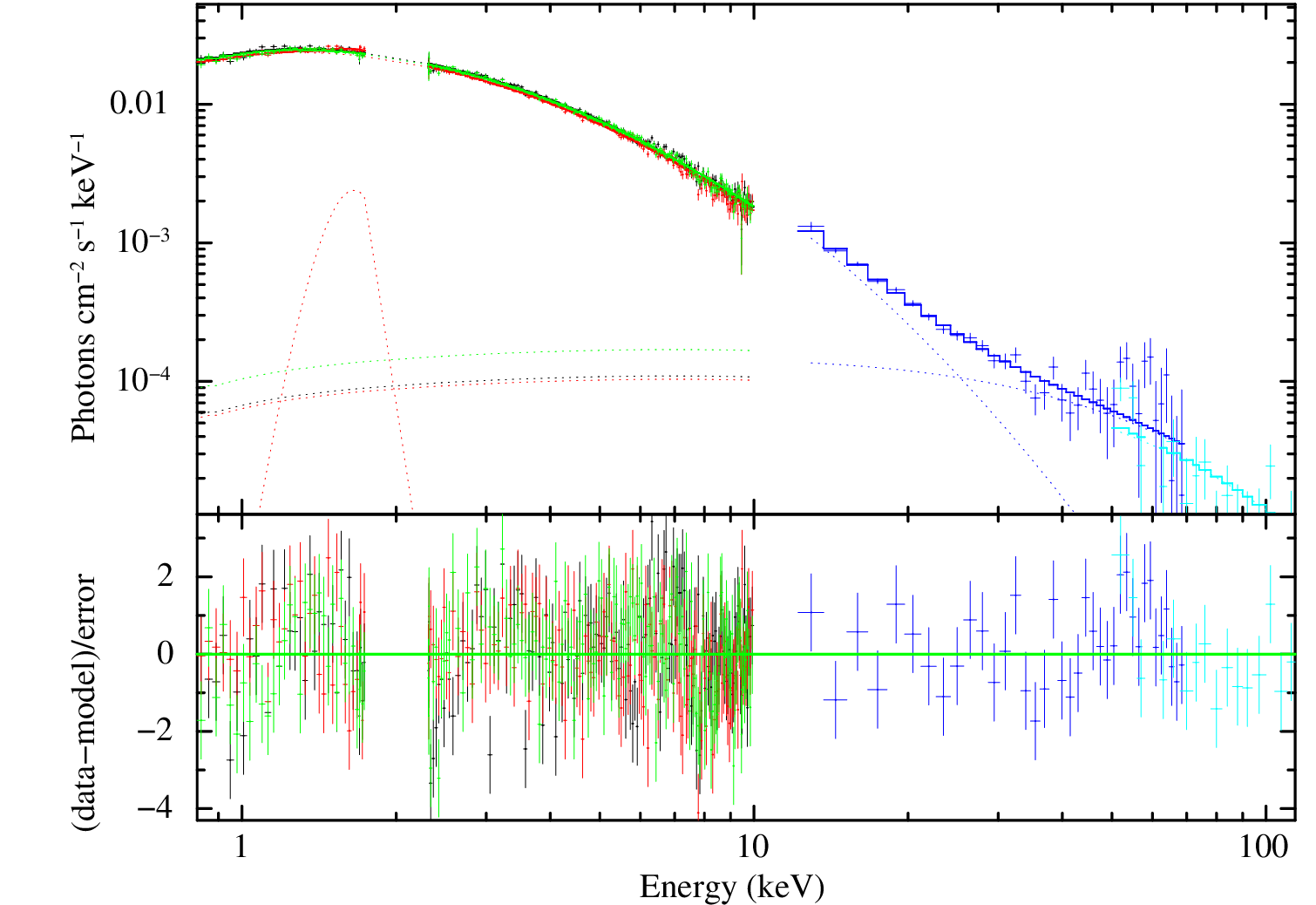}

\caption{Best-fit Unfolded X-ray spectrum of X~Persei at different intensity levels  ('High'- top left; 'Int-2' - top right; 
'Int-1'- bottom left, 'Int-1'- bottom right) constructed 
from XIS (0.5-10 keV), PIN (10-70 keV) and GSO (50-100 keV). The upper panel shows the best fit energy spectrum, 
and the bottom panel shows the residues
of the best fit model without including the CRSF (for 'High' and 'Int-2' spectra).}
\label{spec_int}
\end{figure*}

%\begin{figure*}
% \includegraphics[height=8cm,width=6cm,angle=-90]{crsf_fux.ps}
%\caption{Plot of the cyclotron line energy shown against the flux in the energy band of 10-70 keV.}
%\label{crsf_fux}
%\end{figure*}

\begin{figure*}
 \includegraphics[height=5cm,width=6cm,angle=-90]{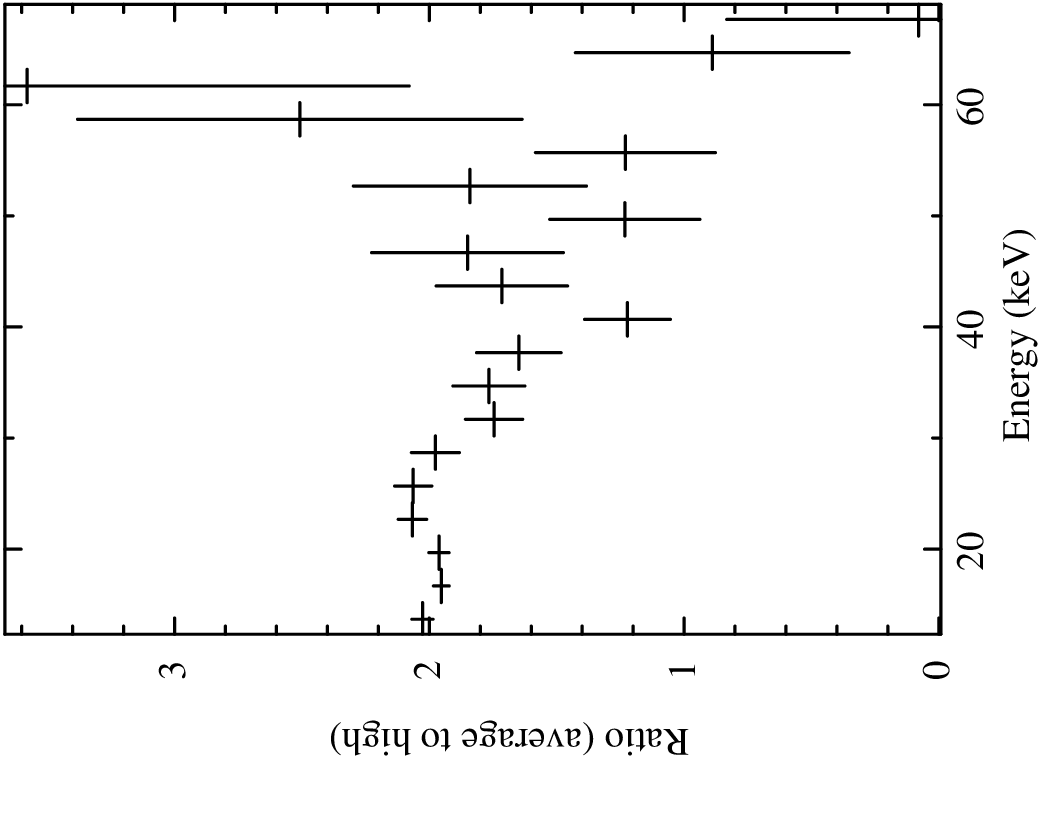}
 \includegraphics[height=5cm,width=6cm,angle=-90]{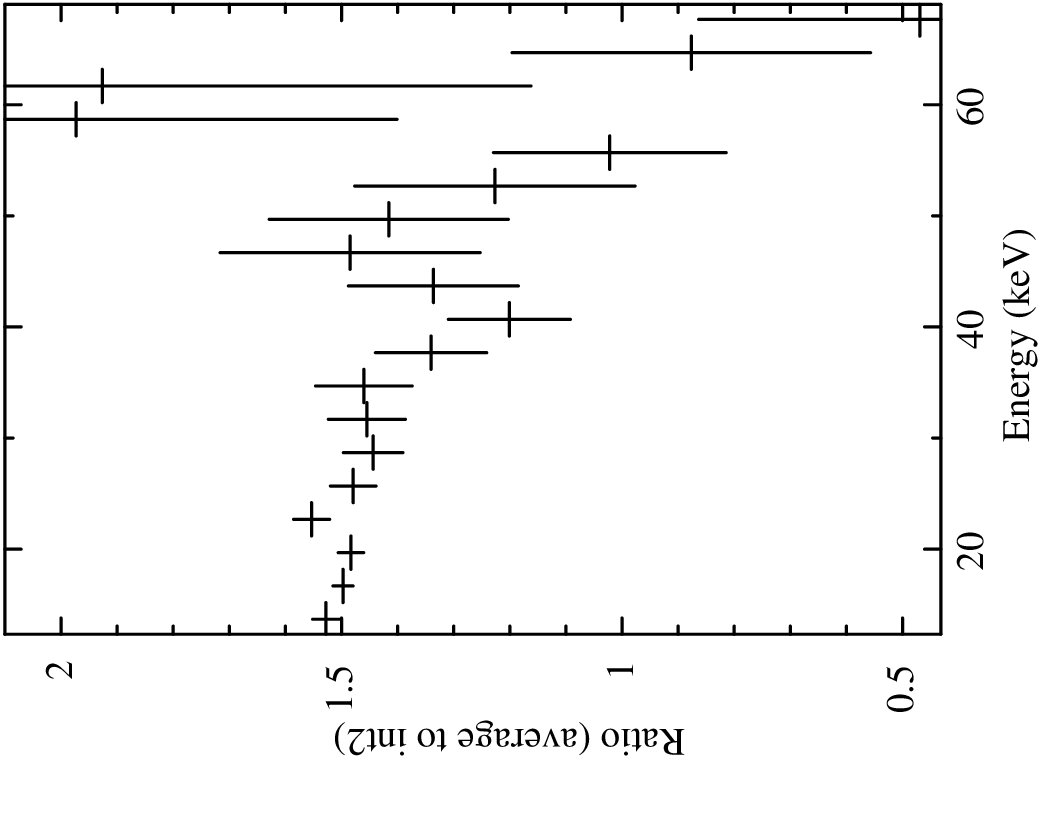}
\caption{Ratio plot of intensity levels ('High' and 'Int2') to average spectra. The dip at $\sim$ 40 keV is evident.}
\label{crsf_ratio}
\end{figure*}

\subsection{Intensity dependence of pulse profiles}
Figure \ref{pp} shows the normalised background subtracted pulse profiles of X~Persei at different intensity levels extracted
from XIS (0.5-10 keV), PIN (10-70 keV) and GSO (50-100 keV) light curves respectively.
The pulse profiles exhibit noticeable evolution with intensity at all energy ranges. For the entire energy band, the profiles become narrower
as the intensity increases. The shoulder like feature between phase range 0.1-0.3 seen at the 'low' and 'int-1' state gradually morphs to a sharp narrow profile. Another important feature is that the corresponding pulse fraction increases. In the XIS energy band the pulse fraction increases from 44 \% to 68 \% between the 'Low' and 'High' intensity levels,
while the corresponding numbers in the PIN energy band are 47 \% and 67 \%. It may be mentioned that \cite{Palombara.et.al.2007} 
reported an increase in pulse fraction with intensity from the {\it RXTE-PCA} data.
Changes in the pulse profiles with intensity have been reported before in several accretion powered pulsars like
Her X-1 \citep{2008A&A...482..907K}, 4U 0115+63 \citep{2007AstL...33..368T} and GX 304-1 \citep{2015A&A...581A.121M}.
The changes can be qualitatively understood due to changes in the height and geometry of the accretion column
with intensity. The results point towards a change in the accretion geometry and a gradual evolution in the 
beaming pattern with increasing intensity with the highest intensity showing the highest beamed pulse profile. Detailed
modelling of the shape of the pulse profiles with intensity requires an accurate description of
the magnetic field geometry, scattering cross sections as well as the energy and intensity dependence of the emission
beam itself, and is beyond the scope of this work.
 . 

\subsection{Intensity dependence of energy spectrum}
We extracted spectrum corresponding to the intensity levels mentioned in section \ref{int}. We started the fits with the 
 best-fit model (obtained with 
'COMPMAG') of the average spectrum to fit the spectra at different intensity levels. 
Like the pulse profiles, the energy spectrum also exhibits dependence with intensity. We noticed the presence of a very weak Fe K$\alpha$
emission line (equivalent width $7.5\pm2.1$ eV), ~in the 'Int-1' intensity state. This might be because, at this intensity level,
the continnum emission is supressed, and yet the statistical quality of the spectrum is enough to detect the line. The electron temperature ($kT$2) of the 'High'
intensity level is higher than in the other spectra, indicating a harder spectrum consistent with the highest pulse fraction of
the corresponding hard X-ray pulse profile (see Fig. \ref{pp}). A correlation between the hardness ratio and intensity, with indication of 
increasing hardness with intensity has been
reported previously in the {\it XMM-Newton} band $<$ 15 keV \citep{2007A&A...474..137L}. The harder spectrum at 'High' intensity may indicate an 
increase in Comptonization in the accretion column. 
The parameter $kT$2 however only traces the thermal Comptonization component. The change in the BMC component, thought to be dominant 
in X~Persei cannot be probed. This is because $\beta$ cannot be 
left free in the fits, due to statistical limitations. 
The errors bar of the parameter $kT$2 is however high due to the low exposure time of
the spectrum, and therefore the increase in $kT$2 in not statistically significant. The other continnum parameters do not show noticeable evolution with intensity.

The most significant changes are dip like features present in the spectrum of the 'High' and 'Int-2' intensities at $\sim$ 40 keV,
reminiscent of a CRSF. Both the residuals and the corresponding $\chi^{2}$ improved after addition of a CRSF feature ('cyclabs'
model in {\it XSPEC}).
 For the 'Int-2' spectrum, additional residuals, reminiscent of another feature was noticed at $\sim$ 29 keV. Adding another
CRSF improved the $\chi^{2}$ further. The width of this feature could not be constrained and was hence frozen at 3 keV.
The total improvement in $\chi^{2}$ for the 'High' and 'Int-2' intensity levels was 20 and 18 after adding the CRSF 
(1 for 'High' and 2 for 'Int-2'
) for 2 d.of and 4 d.o.f respectively. In order to verfiy whether this was a model independent feature, we
checked the spectral fits for the intensity states, using the 'newhcut' spectral model. Residuals are seen at the same energies for the 'High'
and 'Int-2' spectra. However, the dips are much wider and deeper. As discussed in section \ref{avspec}, this difference might arise due to insufficient continnum
modeling of the continnum spectrum with the 'newchut' spectral model. To further verify the presence of the CRSF in a model independent manner
we extracted the ratio of the spectrum of the 'High' and 'Int-2' levels, respectively divided by the average spectrum. 
Figure \ref{crsf_ratio}
shows the ratio plots, with the dip at $\sim$ 40 keV clearly visible. A detailed discussion on statistical significance of the CRSF is included
in the next section.
Figure \ref{spec_int} shows the unfolded spectrum of X Persei (showing the model components) at different intensity levels with and without inclusion of the CRSF feature
for the 'High' and 'int-2' intensities. The best-fit parameters are tabulated in table \ref{avg_spec_params_intensity}, including the CRSFs for 
intensity level 'High' and 'Int-2'.
 
\subsection{CRSF and its detection significance}

Computing statistical significance of CRSFs needs to be handled with care, as this is a multiplicative component modifying the underlying
continnum iself. In order to access whether the improvement in $\chi^{2}$ is significant, the standard F-test cannot be used which is suitable
only for additive components \citep{2002ApJ...571..545P}. Therefore we used two independent tests, to access the significance of CRSFs in the 'High' and 'Int-2' intensity
spectra.

At first we performed a numerical evaluation of the significance of the CRSF by performing Monte-Carlo simulations. This was done with the 
{\it XSPEC} script {\it lrt} by simulating large number of datasets (3001) with the two component 'COMPMAG' model without the CRSF, 
modified with statistical (Poisson) noise. At each iteration the simulated datasets were fitted with the continuum model with and without the
CRSF and the statistics were compared. For the 'High' spectrum, we found that in 824 out of 3001 cases, the simulated $\chi^{2}$ was as high as the 
observed one. This gave a PCI of 27\%. In case of the 'Int-2' spectrum with a similar analysis, the obtained result was 34\%.
Considering the high PCI values in both cases, the CRSF at $\sim$ 40 keV cannot be considered statistically
significant. Performing the simulation with 2 CRSFs instead of one, for the 'Int-2' spectrum gave only slightly better results and cannot be considered
statistically significant.

The evaluation of the statistical significance of a CRSF, which is often a weak absorption feature against the continuum is
recommended to be
supported by other tests, especially one which is sensitive to the structure of the residuals, in order to distinguish
a random fluctulation from a systematic structure \citep{orlandini,2011arXiv1106.2067S}. In the spectral fits of the 'High' and 'Int-2'
spectra, the errors dominate as is evident from the best-fit parameters in table \ref{avg_spec_params_intensity}.
Therefore, we performed a run-test \citep{barlow} on the residuals between 30--50 keV (range in which CRSF is detected)
for the 'High' spectrum, in order to test for the null
hypothesis of the randomness of the residuals of the fit in that energy range. The number of data points (N$_{t}$) was 14, with 3 
points above zero
(N$_{+}$), and 11 below zero (N$_{-}$). This gave a probability of obtaining
N$_{r} \le$ 3 (N$_{r}$ is the number of runs) equals 0.009 (0.9\%). In the case of 'Int-2' spectrum in the same energy range,
N$_{t}$ was 14, with 5 N$_{+}$, and 9 N$_{-}$, which resulted in probability of obtaining
N$_{r} \le$ 3 equals 0.003 (0.3\%). The obtained values
strongly indicated that the residuals are
not due to random fluctulations, and have a systematic structure, indicating the need to add the CRSF component to the spectra. For the 'Int-2'
spectrum, we also performed the test between 25-41 keV, to look for the additional absorption feature at 29 keV.
In this case, N$_{t}$ was 11, with 3
N$_{+}$, and 8 N$_{-}$. This gave a probability of obtaining
N$_{r} \le$ 3 equals 0.01 or a 1\% probability of the structure being due to random fluctulations. The low values obtained in all three
cases provide statistical evidence
that the residual patterns are not due to random fluctulations. 

Although the detection of the CRSF in the high intensity levels of X~Persei is not very statistically significant
in terms of the delta $\chi^{2}$ and the comparison of the $\chi^{2}$ statistics with and without inclusion of the CRSF in the
spectral model, there is strong indication that the feature is indeed present in the spectrum. This is supported from the results of
the run-test
and the model independent verification of the feature. Future deep observations will ascertain this further.

%\begin{figure}
%\vspace{-0.9cm}
% \hspace{-0.9cm}
%%\includegraphics[scale=0.20]{plots_talk/xis.pdf}
% \hspace{-0.5cm}
%%  \includegraphics[scale=0.20]{plots_talk/gso.pdf}
%   \hspace{-0.5cm}
%%\includegraphics[scale=0.20]{plots_talk/pin.pdf}
% \vspace{-0.6cm}
% \caption{Intensity resolved pulse profiles constructed using XIS (left:0.5-10.0 keV), 
% PIN (centre: 10-70 keV), and GSO (right: 50-100 keV). The pulse profiles in high, 
% int-2, int-1 and low intensity states are shown in black, blue, green and red respectively.}
% \label{avg_spec1}
%\end{figure}
%%%%%%%
%\begin{figure}
%\centering
%\vspace{-0.9cm}
% \hspace{-0.9cm}
%%\includegraphics[scale=0.30]{plots_talk/crsf-corr.pdf}
% \vspace{-0.6cm}
% \caption{CRSF centroid energy plotted against the observed intensity states. A positive correlation is observed between the two.}
% \label{avg_spec1}
%\end{figure}

\section{Discussion}

We have performed a comprehensive analysis of X~Persei using a long \emph{Suzaku} observation where we have probed several timing and spectral aspects 
of the source in detail.  The broadband sensitive instruments onboard \emph{Suzaku} have allowed us to model the continuum spectra robustly
and make comparison 
amongst several spectral models. We have found a two component model suitable for the hard X-ray spectrum extending up to 100 keV as in previous
works. There is also strong evidence that BMC dominates the Comptonization process as expected for low luminosity accretion powered
pulsars with a hard continuum spectrum.
The most intriguing aspect of this work is that we have identified for the first time, associated with changes in the pulse 
profile, corresponding indications of changes in the continuum spectra with varying intensity levels, pointing to changes in the accretion 
geometry. We have also found evidence of a CRSF at $\sim$ 40 keV at the highest intensity levels, reinstating the magnetic field strength
of the neutron star.

\subsection{Broadband continuum modelling}
Several previous works have attempted to model the energy spectrum of X~Persei, the most notable being \cite{Salvo.et.al.1998} (using \textit{BeppoSAX} observation), 
\cite{Coburn.et.al.2001} (using \textit{RXTE} observation) and more recently \cite{Doro2012} (using \textit{INTEGRAL} observation). However these results were based on 
instruments which were sensitive over different energy ranges and were unable to constrain either the low energy part of the spectrum (in case of \textit{RXTE} and 
\textit{INTEGRAL}) or the high energy part (in case of \textit{BeppoSAX}) well. X~Persei being a low luminosity source with the energy spectrum ranging below 1 keV to 
above 100 keV, the instruments onboard \textit{Suzaku} were ideal in this regard. We have tested all the continuum models presently avaiable
 to fit the broadband spectrum of accretion powered pulsars. In all 
the cases we required a two component model separately for the low and high energy part of the spectrum as have also been reported in 
previous works. 
We find that a powerlaw component with a smooth exponential rollover provides the best-fit both in terms of reduced $\chi^{2}$
and 
realistic values for the continuum parameters. The continuum model parameters we obtain are consistent with \cite{Salvo.et.al.1998} who use the same model to describe 
the \textit{BeppoSAX} spectrum. Our results are also consistent with \cite{Coburn.et.al.2001} who use a blackbody component to model the low energy part and a powerlaw component 
with an exponential rollover for the high energy ranges of the spectrum. This is because, while the parameters of the hard powerlaw component are similar, we note that 
\textit{RXTE}-PCA being sensitive only $>$ 3 keV cannot differentiate between a 1.7 keV blackbody component (that they obtained) from a 
$\Gamma$ $\sim$ 0.3 powerlaw 
having a rollover at $\sim$ 3.5 keV (current work). The fit with the 'newhcut' continuum model however indicated
a broad absorption feature at $\sim$ 30 keV in the residuals which can be identified with the CRSF previously reported in the source. 
Spectral fit with the analytical model 'COMPMAG' also provides an acceptable fit, and allows constraining physical parameters related to
the accretion on the pulsar. The absorption feature at 30 keV is not required by this model indicating that it might be an effect due to insufficient continuum modelling with the phenomenological models, and needs
to be treated with caution.

The spectral fit with 'COMPMAG' model
fitted with simple assumptions suitable for low luminosity accreting pulsars
 provides a direct insight into the process of spectral formation, and the conditions of the accretion column.
The best-fit average spectrum indicated a high value of $\beta$ indicating that BMC dominates the Comptonization process
in X~Persei.

The broadband unabsorbed bolometric luminosity of X~Persei (0.3-70 keV) is  $L_{x}$ =1.6$\times$10$^{35}$\,erg s$^{-1}$. 
In accretion powered pulsars with luminosity 
$L_{x}$ $<$10$^{37}$\,erg s$^{-1}$, it is widely believed that the radiation is mainly generated from the hotspots generated by the infalling plasma which decelerates 
to the surface of the neutron star due to Coulomb interactions forming an accretion mound \citep{burnard1991}.
In such sources, BMC is
believed to play an important role. Further, this model has been particularly preferred to fit the spectrum of steep sources like X~Persei 
\citep{wolff2007,wolff2005}. This is in agreement with our obtained results. The hard power-law tail in the PIN energy range wih a 
rollover at  energies much higher than expected for accreting pulsars ($\sim$ 56 keV),
is in further support of this.

\subsection{Intensity levels}

 We report the presence of large flux variations in the observation and divided the duration into four different intensity levels: a) High b)
Intensity 2 c) Intensity 1 and d) 
Low ordered by decreasing count rates where the difference between the highest and the lowest intensity levels is factor of $\sim$ 5. 
To investigate whether there are distinct timing and spectral characteristics
of the source at different intensity levels, we extracted pulse profiles in the low and high energy band as well as the broadband 
spectrum corresponding to these intensity levels. We 
confirm that there are indeed associated changes in the pulse profile, the corresponding spectrum, with
 the source flux.
For all the energy ranges we see that the pulse profiles become narrower with increasing intensity with an increase in pulse fraction from 
44 \% to 68 \% at the highest 
intensity level. There is also an indication that the highest intensity level ('High')
has a harder spectrum. The highest intensity levels ('High' and 'Int-2') also indicate the presence of a CRSF at $\sim$ 40 keV.
The changes in the pulse profile and the energy spectrum reflects a change in the beam 
function and a change in the conditions of the accretion column with intensity. 

In accretion powered X-ray pulsars the 'critical luminosity' $L_{\mathrm{crit}}$ divides two regimes of accretion 
into the super and sub-critical regime \citep{becker2012}. The equation is given by 

\begin{eqnarray}
	L_{\rm crit} &=& 1.49 \times 10^{37}{\rm erg\,s}^{-1} \left( \frac{\Lambda}{0.1} \right)^{-7/5} w^{-28/15} \nonumber \\
	&&\times \left( \frac{M}{1.4{\rm\,M}_{\odot}} \right)^{29/30} \left( \frac{R}{10{\rm\,km}} \right)^{1/10} \left( \frac{B_{\rm surf}}{10^{12}{\rm\,G}} \right)^{16/15}
	\label{eqn:lcrit}
\end{eqnarray}
Here $R$, $M$, and $B$ are, the radius, mass, and surface
magnetic field strength (B in terms of $\sim 10^{12}$ G) of the neutron star.
The parameter  $\Lambda = 0.1$ approximates the case of disk accretion, and $\Lambda
= 1.0$ is suitable for wind accretors, 
although it is difficult to determine the parameter precisely for an assumption of spherical accretion.
$w = 1$ on the other hand, characterizes the shape of the
photon spectrum inside the column (where there is assumed to be a mean photon
energy of $\bar{E} = wkT_{\rm eff}$), and $\Lambda$ characterizes the mode of
accretion. For X~Persei it is reasonable to assume the value of 
$\Lambda$ for wind accretors \textit{i.e.} 1. Therefore we obtain $L_{\rm crit}=2.1\times10^{36}$\,erg $s^{-1}$ assuming $B=3.4\times10^{12}$\,G 
(obtained from the 
CRSF centroid energy of $\sim$ 40 keV), and standard values for the mass and radius of the neutron star.
This is above the bolometric luminosity of $1.4\times10^{35}$\,erg s$^{-1}$  of the source, which indicates the source is in the 
sub-critical regime with the emission dominating from the accretion mound. In this regime, the characteristic emission height, from where most of the radiation escapes 
decreases with increasing luminosity. So at higher intensity if the accretion mound is pressed closer to the neutron star surface owing to higher ram pressure of the 
infalling material, one might be probing the hotter base of the accretion mound with higher fraction of higher energy photons generated due to Comptonization, which gives rise to a harder and more beamed emission.
Our results are also consistent with \cite{postnov2015} which investigated the behaviour of the spectral hardness ratio
as a function of X-ray luminosity in a sample of X-ray pulsars at an energy range $<$ 12 keV (using {\it RXTE-ASM} data). 
The authors observed an increase in hardness ratio with increasing luminosity
for sources in the sub-critical regime. The increasing hardness with luminosity in X~Persei are observed at energies $\le$ 110 keV 
(The upper limit for the spectral fitting band), and indicates that the postulate is also valid in a broad energy band.

\subsection{Cyclotron line}
A very interesting outcome of this work is the indication of the presence of the CRSF at the high intensity levels of the source.
The presence of the CRSF has been long debated in X~Persei. It's presence was first reported by \cite{Coburn.et.al.2001} 
at $28.6^{+1.5}_{-1.7}$ keV, but was not confirmed later from the \textit{BeppoSAX} and \textit{INTEGRAL} observations.
We report a strong evidence for the presence of the line at $38.6 \pm 4.4$ keV $40.8 \pm 2.3$ keV, for the 'High' and 'Int-2' intensity levels respectively. We however do not find a conclusive evidence 
for the presence of the line in the average spectrum. The detection of the CRSF at a higher
centroid energy than in the \textit{RXTE} observations might be related to
the difference of intensity between the two by a factor of $\sim$ 3. The centroid energy of the CRSF 
varies with intensity. A positive correlation with luminosity has been detected in several other low luminosity sources like GX 304-1 
and Her X-1 \citep{staubert2007,yamamoto2011,klochkov2012}. 
The measured luminosity of X~Persei being sub-critical, the source perfectly fits in this group. In the sub-critical accretion regime, 
the CRSF energy increases at higher
intensities as the accretion mound is pushed more towards the surface of the neutron star with increasing ram pressure thereby sampling 
higher magnetic field 
(assuming a dipolar geometry for the magnetic field). The obtained result also points at another important aspect that the CRSF in X~Persei
is preferably detected at higher intensities. 
This was already hinted at \cite{Coburn.et.al.2001} to explain why the CRSF feature was not detected in the \textit{BeppoSAX} observation. 
In table \ref{flux} we list 
the 1--10 keV flux obtained from different observations of the source using different instruments.
The results indicate that the CRSF is detected at higher intensities. 
In fact the \textit{Suzaku} observation seems to have caught the source at the highest average intensity observed till now. The only other report is from the {\it XMM-Newton} observation of the source, 
which was reported to be at the same luminosity \citep{Palombara.et.al.2007}. However the {\it XMM} band does not allow the detection of the CRSF.

The identification of the line centroid energy is one of the surest method to estimate the magnetic field strength of the emitting region in the neutron star. This is given by the ``12-B-12'' rule providing an approximate relation
between the observed CRSF energy and the magnetic field strength in the
scattering region:
\begin{equation}
	E_{\mathrm{cyc}} = \frac{11.57\,\mathrm{keV}}{1+z} \times B_{12}
	\label{eqn:12b12}
\end{equation}
where $B_{12}$ is the magnetic field in units of $10^{12}$\,G and $z \sim 0.15$
is the gravitational redshift in the scattering region for standard neutron star parameters. 
This implies the magnetic field strength of the neutron star is
$B=3.4\times10^{12}$\,G, assuming the scattering region lies close to the neutron star surface. 

An intriguing aspect of the CRSF in X~Persei is that the residuals after fitting the continuum model in the 'Int-2' level spectrum
shows two dip like structures, one at $\sim$ 29 keV and another at $\sim$ 40 keV. This is an indication of 
the CRSF being non-Gaussian with a more complex shape. This aspect has been discussed before by \citep[]{pott2005,nakajima2010} and
\citet{Mushtukov2015} for the accreting pulsar V 0332+53 where a similar effect has been observed. A complicated non Gaussian CRSF shape has been long predicted by 
theory \cite{schonherr2007} but has not been confirmed convincingly from observational evidences, except for one source V0332+53. X~Persei will only be the second 
source in this regard. This effect is also expected to be more prominent at lower luminosities as explained by several theories. \cite{bhattacharya2011} predicted 
complicated CRSF shapes in the spectra of low luminosity pulsars. This is due to the distortion of the dipolar magnetic field near the base of the accretion mound due to 
the effect of local instabilities. \cite{Mushtukov2015} also predicted a similar effect for sub-critical sources  due to the variation of the doppler boost with 
changing luminosity as discussed before. The observed CRSF in the low luminosity sub-critical source X~Persei is consistent with these predictions. We encourage 
further deep observations of this source with sensitive instruments having high energy resolution to probe this aspect deeper. 

\begin{table*}[h!]
\caption{Table comparing measured flux of X~Persei using different instruments}
\begin{tabular}{c c c }
\hline \hline

Instrument & Flux & CRSF detected \\
--  & 1--10 keV   &  ---\\
\hline
\textit{BeppoSAX} & 1.7 & NO \\
\textit{RXTE}  & 2.1  & YES \\
\textit{INTEGRAL} & 1.9 & NO\\
\textit{Suzaku (average}) & 6.6 & NO \\
\textit{Suzaku} ('Int-2' intensity & 9.0 & YES \\
\textit{Suzaku}('High' intensity) & 11.7 & YES \\
\hline
\end{tabular}\\
\label{flux}
$^{a}$ Flux is in units of $10^{-10}$ ${\rm ~erg~cm^{-2}~s^{-1}}$\\
\end{table*}

\section{Summary}
We presented a detailed broadband timing and spectral study of X~Persei performed using a \emph{Suzaku} observation. 
By testing different continuum 
models, we have obtained the representative broadband X-ray spectrum of X~Persei.
The spectral fits indicate that BMC 
is a dominating factor for the Comtonization of the seed photons at the base of the accretion column;
both indicated by the requirement of a high $\beta$ in the 'COMPMAG' model
and a hard powerlaw tail like feature with an exponential rollover at $\sim$ 56 keV obtained with the 'newhcut' model
. For the first time, we have 
identified different intensity levels of the source with distinct changes in the pulsation characteristics and the energy spectrum. 
All of these indicate a change in the accretion  geometry of X Persei with varying intensity. 
We have also found evidence of a CRSF at $\sim$ 40 keV
at higher intensities. This may be related to the fact that the CRSF is preferably detected at higher intensities, and may also indicate a correlation
of the line centroid energy with intensity. 

\section*{Acknowledgments}
This work has made use of data obtained from the High Energy Astrophysics Science Archive 
Research Center (HEASARC), provided by the NASA Goddard Space Flight Center. CM acknowledges the hospitality from
Nordita, KTH Royal Institute of Technology and Stockholm University for hosting her for a week, where most of the work
for the paper was done, and very fruitful discussions with Ruben Farinelli for help with the COMPMAG and COMPMAG2 models.

\bibliographystyle{mn2e}
\bibliography{ms}

\end{document}